# Experimental Observation and Description of Bandgaps Opening in Chiral Phononic Crystals by Analogy with Thomson scattering

Wei Ding[1], Tianning Chen[1], Chen Chen[1], Dimitrios Chronopoulos[2], Badreddine Assouar[3], Jian Zhu[1,4]*

[1]*School of Mechanical Engineering and State Key Laboratory of Strength & Vibration of Mechanical Structures, Xi'an Jiaotong University, Xi'an, Shaanxi 710049, P.R. China*
[2]*Department of Mechanical Engineering & Mechatronic System Dynamics (LMSD), KU Leuven, 9000, Belgium*
[3]*Université de Lorraine, CNRS, Institut Jean Lamour, F-54000 Nancy, France*
[4]*School of Mechanical Engineering and State Key Lab of Digital Manufacturing Equipment & Technology, Huazhong University of Science and Technology, Wuhan, Hubei, 430074, P. R. China*

Chiral phononic crystals provide unique properties not offered by conventional phononic material based on Bragg scattering and local resonance. However, it is insufficient to only consider the inertial amplification effect in chiral phononic crystals. Here, we theoretically and experimentally introduce the analogy with Thomson scattering to characterize the bandgap phenomena in chiral phononic crystals. Two phononic structures are proposed and discussed, one with translation-rotation coupling and another with translation-translation coupling. The two lattices are different in appearance but have similar bandgaps. Thomson scattering in electromagnetic waves was drawn on to describe the coupling motion of the unit cells. We evidence that the bandgap generation is essentially based on the analogous Thomson scattering aiming to achieve an anti-phase superposition of the waves in the same polarization mode. This finding sheds new light on the physics of the elastodynamic wave manipulation in chiral phononic crystals and opens a remarkable route for their pragmatic implementation.

Phononic crystals (PnCs) are periodical artificial structural materials with the capability of flexible manipulation of acoustic and elastic waves, which has received much attention[1-3]. The remarkable feature, i.e. the bandgap, provides a powerful ability to control the wave propagation, such as the design of the non-reciprocal device [4-6] and the recent valley state formation with Dirac points degeneracy operation in its bandgap [7,8]. The mainstream formation mechanisms of bandgap include Bragg scattering [9] and local resonance [10]. The local resonant-based PnCs can control the waves at or near the resonant frequency of the vibrator, therefore resulting in a narrow band [11-14], and the Bragg scattering bandgap exhibits the continuity and deep energy suppression but the wavelength corresponding to the center frequency of the bandgap is in the same order of the magnitude of the lattice constant [15,16], or the deep subwaveleghth bandgap in a lattice with extreme connection or out-of-balance size matching between the matrix and the scatterer [17-19].

The introduction of chirality in PnCs or metamaterials provides additional degrees of freedom in wave-matter interactions tailoring and manipulation. A variety of unusual physical properties and exotic functions, including the compression-torsion effect [20-22], extreme softening and compression-induced-twisting behavior [23,24], tunable Poisson's ratio and stiffness [25-27], are obtained in a kind of chiral compression-torsion structures which cannot be offered by conventional materials or PnCs. In particular, the chirality elements in PnCs will produce inertial amplification effect, which can be used to generate a wide bandgap at a low starting frequency [28-32] to elude the shortcomings of local resonance and Bragg scattering.

As a bandgap mechanism independent of local resonance and Bragg scattering, inertial amplification method requests two crucial conditions from the perspective of the inertial matrix [28,33]. In addition to the fact that the dynamic inertia must be larger than the static inertia, i.e. inertial amplification, the coupling between the lumped masses is required, i.e. inertial coupling [33]. Although, in classical inertial amplification-based systems, it is feasible to consider only inertial amplification [34-37]. However, for the chiral compression-torsion structures [31,32,38,39], despite the inertial amplification effect in the chiral subunit cell (see Supplementary S1), inappropriate geometrical configurations won't form the bandgap [32]. In other words, considering the inertial amplification of a subunit cell only is not enough to capture the underlying physics and the bandgap of the compression-torsion coupling-induced PnCs [31,32]. Furthermore, the concepts of inertial amplification and inertial coupling are abstract from the perspective of the wave propagation, leading to extensive designs based primarily on classical inertial amplification models, and thus limiting the other considerations beyond classical inertial amplification-based design strategies.

Here, we theoretically and experimentally investigate the wave phenomena and bandgap generation in compression-torsion coupled PnCs to concretize the bandgap mechanism. An incident wave polarizing in one mode passing the chiral subunit cells will be decomposed into the outgoing waves with two polarizations, the one in translation and the other in rotation. The decomposition of the polarisation can be considered as a splitting process of the incident waves, which can be an analogy to Thomson scattering in electromagnetic waves [40,41]. It demonstrates that a minimum of twice Thomson scattering and the outgoing waves vibrating in identical mode have opposite phases after the second Thomson scattering are required to generate a bandgap. The findings are verified in a kind of chiral PnCs with translation-rotation coupling. For the universality of the underlying physics, another non-chiral lattice with the translation-translation coupling is proposed to demonstrate that this bandgap is not unique to chiral lattices but available in other lattices with Thomson scattering effects.

Figure 1 shows the schematics of the unit cells and calculated band structures. The lattices in Fig. 1(a) and Fig. 1(b) are unit cells with translation-rotation coupling, and in Fig. 1(c) and Fig. 1(d) are unit cells with translation-translation coupling. Each type of lattice consists of two subunit cells. As depicted in Figs. 1(a) and 1(b), the arrayed translation-rotation coupling unit cell named ATR consists of two subunit cells **I**. The mirrored translation-rotation coupling unit cell called MTR consists of subunit cells **I** and **II**. The right panels in Figs. 1(a)-(b) represent the corresponding band structure calculated by utilizing finite-element method based software Comsol Multiphysics. In terms of the bandgap, there is an extensive bandgap in MTR but not available in ATR.

In fact, the research on such bandgaps in MTR can be traced back to Bergamini's work in 2019 [32], where the bandgap mechanism is interpreted as inertial amplification [28]. However, as illustrated in Eq. (S58) and Eq. (S63) in Supplementary S2, we found that the inertial matrixes of MTR and ATR have the inertial amplification effect, but the bandgap only exists in MTR. Therefore, inertial amplification lacks sufficient universality for a mechanism like Bragg scattering and local resonance.

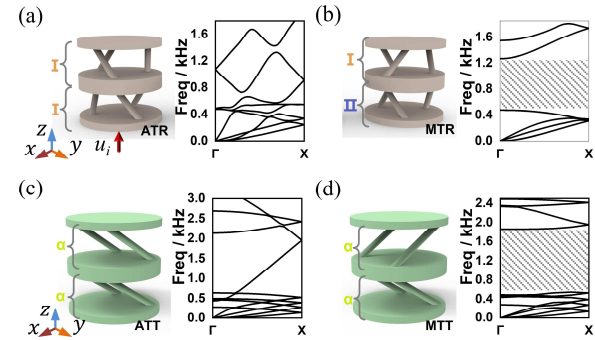

Fig. 1 The lattices and the band structures. (a) Arrayed translation-rotation coupling unit cell (ATR); (b) Mirrored translation-rotation coupling unit cell (MTR); (c) Arrayed translation- translation coupling unit cell (ATT); (d) Mirrored translation-translation coupling unit cell (MTT). The symbols "**I**" and "**II**" represent two different subunit cells of translation-rotation coupling lattices (More details of the geometry can refer to Fig. S1.). Symbol "**I**" is left-handed, and symbol "**II**" is right-handed. The symbol "**α**" is the subunit cell of translation-rotation coupling lattices, and its geometry is illustrated in Fig. S6.). The right panels in (a)-(d) represent the corresponding band structure with the bandgaps highlighted by gray areas.

As illustrated in the analysis of the subunit cell **I** (more details in Supplementary S1), the chiral effect essentially achieves the motion coupling, and thus exhibits the function of the inertial amplification, but this is not sufficient to form a bandgap. Due to the ambiguous contribution of chirality to the bandgap generation, one may easily attribute the underlying physics of the bandgap in Fig. 1(b) to the chirality. For instance, in some specific functional structures, chirality can indeed bring novel phenomena [42], such as negative Poisson's ratio [43], high structural damping [44], and spin mechanical metastructures [20]. Some of these aforementioned properties have been realized in other non-chiral structures [18,45,46]. Therefore, chirality is one of the effective ways but not the exclusive way to achieve the desired function.

The designed non-chiral unit cells with translation-translation coupling consist of two subunit cells *a* are demonstrated in Fig. 1(c) and Fig. 1(d). The arrayed and mirrored translation-translation coupling unit cell with central and rotational symmetry in α is named ATT and MTT. The right panels in Figs. 1(c)-(d) represent the calculated band structure. From the perspective of the bandgap, there is a phenomenon similar to that between MTR and ATR in Figs. 1(a)-1(b), i.e., MTT has the bandgap that is not in ATT, which is an encouraging result as it confirms that the bandgap possessed in MTR is not exclusive to chiral lattices.

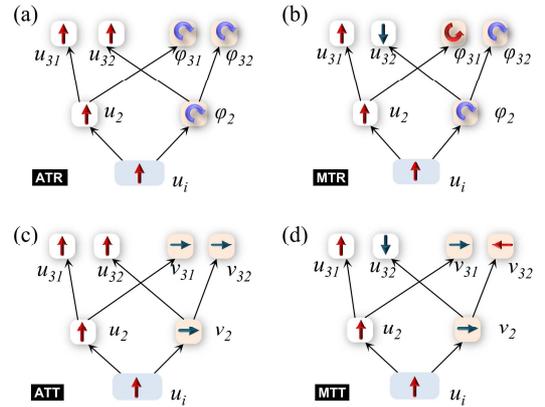

Fig. 2. The decomposition of the movement for (a) ATR, (b) MTR, (c) ATT, and (d) MTT.

To elucidate the mechanism of bandgap generation, Fig. 2 shows the initial vibration orientations of the oscillators when the wave propagates in four types of unit cells shown in Fig. 1. For the sake of simplicity, two assumptions are made during the analysis. The $z$-axis rotational freedom of $m_1$ is restricted; the incident wave vibrates sinusoidally in translational form, and the initial direction of vibration is along +z axis with an initial phase of zero (refer to the coordinate system shown in Fig. 1). As shown in Fig. 2(a) or Fig. 2(b), when the incident wave passes through the first subunit cell, the translation $u_i$ will split into the translation $u_2$ and rotation $\varphi_2$.

Actually, the splitting process is similar to the Thomson scattering. During the Thomson scattering, the plane light forces the electrons to produce forced vibration; the vibration of electrons will create electromagnetic waves with the same frequency as the incident wave [40]. In terms of the function, the subunit cell can be considered as the electron in Thomson scattering. In terms of the scattering process, the forced vibration of the electrons driven by the incident electromagnetic wave is similar to the vibration of the chiral subunit cell operated by the incident elastic wave. In terms of the results, translational and rotational coupling in chiral subunit cells resembles the interaction of the new electric field and the new magnetic field generated by the vibration of the electron in classical Thomson scattering. The change of the wavefront in classical Thomson scattering is a superimposing effect of scattered waves propagating in multiple directions [47,48]. However, limited by the layout of the chiral subunit cell, the scattered waves cannot diverge to infinite space like the classical Thomson scattering but can only propagate in the direction of the periodic structure. These similarities enable us to draw an analogy with Thomson scattering to clarify the process of wave propagation in chiral subunit cells. Therefore, for simplicity, the wave propagation process in the chiral subunit cell is named Thomson scattering.

Furthermore, the process that these scattered waves ($u_2$ and $\varphi_2$) pass through the second subunit cell is equivalent to undergoing a second Thomson scattering. The distinction from the first is that the second scattering has two incident waves ($u_2$ and $\varphi_2$). Consequently, after the second scattering, it will produce four scattered waves, i.e., $u_{31}$, $u_{32}$, $\varphi_{31}$, and $\varphi_{32}$. Among them, $u_{31}$ and $u_{32}$ vibrate in translation, but $\varphi_{31}$ and $\varphi_{32}$ vibrate in rotation.

In ATR, by the first scattering, the initial direction of the scattered waves $u_2$ and $\varphi_2$ are $+z$ and clockwise around the $+z$ axis (from the +z-axis perspective), respectively. After the second scattering, regarding $u_2$ as the incident wave, the initial direction of the scattered wave $u_{31}$ is +z, and the initial direction of the scattered wave $\varphi_{31}$ is in the clockwise direction. Meanwhile, regarding $\varphi_2$ as the incident wave, the initial direction of $u_{32}$ is +z, and the initial direction of $\varphi_{32}$ is also in the clockwise direction. Since the two translations $u_{31}$ and $u_{32}$, and the two rotations $\varphi_{31}$ and $\varphi_{32}$ have the same direction, $u_3$ and $\varphi_3$ of $m_3$ in ATR can be written as Eq. (1) and Eq. (2).

$$u_3 = u_{31} + u_{32}. \tag{1}$$

$$\varphi_3 = \varphi_{31} + \varphi_{32}. \tag{2}$$

Eventually, the absolute angle $\varphi_3$ of $m_3$ in ATT can be determined as

$$\varphi_3 = q(u_i - u_3), \tag{3}$$

where $q$ is the rotational angle of translation per unit (see Supplementary S2 for details).

Conversely, in MTR, the initial direction of the scattered wave $u_{31}$ is +z, and the initial direction of the scattered wave $\varphi_{31}$ is in the clockwise direction; nevertheless, the initial direction of $u_{32}$ is -z, and the initial direction of $\varphi_{32}$ is in the counterclockwise direction. The $u_3$ and $\varphi_3$ of $m_3$ in MTR can be written as

$$u_3 = u_{31} - u_{32} \tag{4}$$

and

$$\varphi_3 = \varphi_{31} - \varphi_{32}. \tag{5}$$

Then, in MTR, the absolute angle $\varphi_3$ can be determined as

$$\varphi_3 = q(u_i + u_3 - 2u_2). \tag{6}$$

The difference in the absolute angle for $m_3$ in ATR and MTR leads to the immense variability of the elements in inertial matrixes $\boldsymbol{M}$ and $\boldsymbol{M}'$ in Eq. (7) (more details in Supplementary S2). In brief, after twice Thomson scattering, the property that the scattered waves vibrating in the same modes have opposite initial vibration directions similar in MTR plays a decisive role in the presence of the bandgap.

$$\boldsymbol{M}\ddot{\boldsymbol{u}} + \boldsymbol{K}\boldsymbol{u} = \boldsymbol{M}'\ddot{u}_o + \boldsymbol{K}'u_o. \tag{7}$$

The splitting process and wave propagating in the translation-translation coupling (ATT and MTT) PnCs can also be an analogy to Thomson scattering, as shown in Fig. 2(c) and Fig. 2(d). In ATT, by the first scattering, the initial directions of the scattered waves $u_2$ and $v_2$ point in +z and +x axis. $u_{31}$, $u_{32}$ and $v_{31}$, $v_{32}$ point to +z and +x directions after the second scattering when $u_2$, $v_2$ are considered as the incident wave. However, in MTT, $u_{31}$ and $u_{32}$ point to +z and -z, $v_{31}$ and $v_{32}$ point to -$x$ and +$x$ directions, respectively. The superposition of these waves in MTT with opposite vibrating directions enables the capability of energy cancellation and generating bandgap, coinciding with the findings in MTR. Therefore, the analogy results confirm the universality of the mechanism proposed in this work.

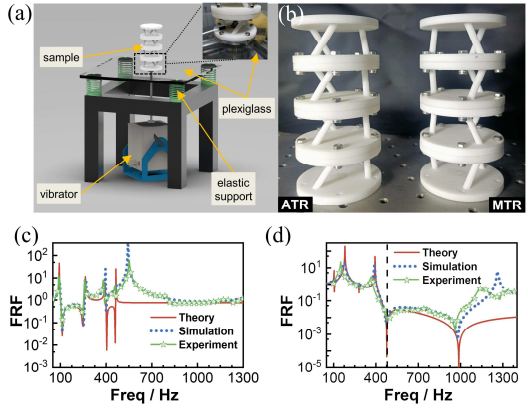

Fig. 3. (a) The schematic of the experimental configuration. The insert in the upper right corner is the detail of the connection between the sample and the plexiglass. (b) Photograph of the investigated samples. ATR sample on the left, and MTR sample on the right. (c) Experimental (green solid with stars), numerical (blue dotted), and analytical (red solid) frequency response functions (FRFs) of ATR. (d) Experimental (green solid with stars), numerical (blue dotted), and analytical (red solid) FRFs of MTR.

To verify the reasonability of the wave phenomena and bandgap feature described by using the analogy of Thomson scattering in Fig. 2, we have built the movement relationship in formulas according to the scattering process. Meanwhile, taking two unit cells in Fig. 3(b) as an example, the theoretical FRFs is carried out (more details of the theory in Supplementary S2). It is clear that, at high frequencies, the transmission ratio of MTR and ATR is constant, but only the MTR exhibits significant attenuation, which allows the excellent potential for generating an ultra-broad bandgap in this lattice. Comparing Fig. 3(d) and Fig. S3 (more details can be found in Fig. S7), it can be seen that the number of the anti-resonant notches will increase as the unit cell periodicity increases, which facilitates enhancing the attenuation of bandgaps. However, for the purpose of exploiting these excellent properties in practice, controlling the resonant modes at the upper boundary of the bandgap at higher frequencies is necessary, as well as a new challenge.

We numerically calculated the FRFs of the MTR and ATR as shown in Figs. 3(c) and 3(d). In the numerical calculation, the rotational freedom of the input disc is restricted to be consistent with the boundary conditions of the theoretical analysis. The strong attenuation is obtained from 400 Hz to 1200 Hz in MTR but not no attenuation in ATR. The theoretical result represented by the solid red line matches well with the simulation result, as indicated by the blue dotted line.

We further experimentally verified the above analysis of the finite periodic structure by measuring the FRFs of the dual unit cells for MTR and ATR. Figure 3(a) shows the schematic of the experimental setup. Figure 3(b) shows the fabricated ATR and MTR samples through photopolymerization-based 3D printing technology (see Supplementary S4). As shown in the insert of Fig. 3(a), the input disc and the plexiglass plate are fixed together by bolts to approximately achieve the constraint condition as the theoretical analysis. The Plexiglas plate is about ten times heavier than the disc (see Supplementary S4 for more details on the simulation and experiment). As illustrated in Figs. 3(c) and 3(d) (the theoretical (red line), simulated (blue dotted line) and experimental results (green line with stars) of FRF), a significant attenuation occurs in MTR, and it does not exist in ATR. In a nutshell, the bandgap does only exist in MTR, not in ATR. Besides, the consistency of experimental, numerical, and theoretical results demonstrates the correctness of the analysis. Also, it verifies the validity of the analogy that the bandgap generation mechanism of this PnC is similar to Thomson scattering.

This analogy can reasonably explain several phenomena from the perspective of underlying physics. It is well known that the Bragg scattering is the elastic collision between the waves and the atoms. The scattered waves in Bragg scattering rely on the heavy atoms to reflect the incident waves and thus to destructive interferences [49]. Thomson scattering depends on the polarization of electrons to generate the divergent outgoing waves in different polarizations [41]. In the Bragg scattering lattice, the propagation of the scattered wave does not depend on the vibration of the scatterer represented by the lumped mass, and thus the scattered waves propagate mainly as reflections [49] with the vibration concentrated on the ligaments [15,50] (please see Supplementary S8). In Thomson scattering, however, the propagation of the scattered waves depends on the polarization of the electrons represented by the entire subunit cell. The orientation of the scattered waves is primarily in the forward direction of the incident wave. As a result, the attenuation in the bandgap is gradient rather than localized on ligaments (please refer to Fig. S7). Because the electrons are much lighter in mass than the atoms, the Thomson scattering will produce a more lightweight unit cell than a Bragg scattering unit cell for the same lattice, stiffness, and bandgap starting frequency [51,52]. In addition, Thomson scattering does not require multiple periodicities, allowing the less periodic structure to reflect the significant attenuation of the bandgap (please refer to Fig. S3 and Fig. S6).

In conclusion, we have theoretically and experimentally demonstrated the wave propagation and the formation mechanism of bandgaps in compression-rotation coupling-induced PnCs. The coupling motion and wave propagating profile can be an analogy to Thomson scattering. The results revealed that several conditions need to be met in the Thomson scattering-based PnCs to generate a bandgap. Firstly, the orthogonal coupling motions are essential for producing the Thomson scattering, which is quantitatively characterized as inertial amplification in equations. Secondly, these initial scattered waves need to undergo a minimum of twice Thomson scattering. Importantly, the secondary scattered waves vibrating in the same mode should have the opposite initial direction of the vibration, which is the cause of the superimposable attenuation, and thus generates the bandgap. Notabely, it is these opposite initial directions produced in the second round of the Thomson scattering, forming a non-diagonal inertial matrix, which is the essence of the inertial coupling. Although the quantitative characterizations demonstrate the antiresonance frequencies in FRFs, the antiresonance frequency is not essential for the bandgap generation (please refer to Supplementary S9). This work revealed that PnCs with inertial amplification only do not have an extensive bandgap similar to MTR and MTT. Chirality is a virtual design element and one of the methods in realizing the Thomson scattering but not an indispensable condition for forming such a bandgap. According to the materialisation of the inertial amplification and inertial coupling carried out in this work, it implies the possibility of coupling other two or more orthogonal modes to create a lower bandgap. The works could shed new light on the physics of the elastodynamic wave manipulation in inertial amplification-induced PnCs and offer an entirely exotic avenue for the further design and investigation of PnCs with remarkable properties, such as the bandgap with lower starting frequency, broadband and extensive attenuations.

This work was financially supported by the National Natural Science Foundation of China (No. 12002258), the State Key Lab of Digital Manufacturing Equipment & Technology of HUST (DMETKF2021014), the Natural Science Foundation of Shaanxi Province (No. 2020JQ-043).

* Corresponding author: jianzhuxj@xjtu.edu.cn